# A bibliometric methodology to unveil territorial inequities in the scientific wealth to combat COVID-19

**Authors:** Giovanni Abramo[1], Ciriaco Andrea D'Angelo[2,1]

**Affiliations**

[1] Laboratory for Studies in Research Evaluation, Institute for System Analysis and Computer Science (IASI-CNR). National Research Council, Rome, Italy. ORCID: 0000-0003-0731-3635 - giovanni.abramo@uniroma2.it

[2] University of Rome "Tor Vergata", Dept of Engineering and Management, Rome, Italy. ORCID: 0000-0002-6977-6611 - dangelo@dii.uniroma2.it

**Corresponding author**

Giovanni Abramo
Consiglio Nazionale delle Ricerche
Istituto di Analisi dei Sistemi e Informatica
Via dei Taurini 19, 00185 Rome - ITALY
+39 06 72597362

**Contributorship:** The authors contributed equally to this work.

**Abstract**

In this paper we develop a methodology to assess the scientific wealth of territories at field level. Our methodology uses a bibliometric approach based on the observation of academic research performance and overall scientific production in each territory. We apply it to assess disparities in the Italian territories in the medical specialties at the front line of the COVID-19 emergency. Italy has been the first among western countries to be severely affected by the onset of the COVID-19 pandemic. The study reveals remarkable inequities across territories, with scientific weaknesses concentrated in the south. Policies for rebalancing the north-south divide should also consider, in addition to tangible assets, the gap in production and availability of quality medical knowledge.

## 1. Introduction

As we write this manuscript, ten weeks have elapsed since the first Italian case of COVID-19 contagion. To date, 203,951 persons have been infected, and 27,682 of these have died.[1] Italy was the first western country to suffer the disastrous force of COVID-19, and to date it is the second country in the world for number of victims, preceded only by the USA. The population has sacrificed one of the founding national values, that of Freedom, so costly in human lives and fresh in the collective memory, given the events for of our country in the 20th century. Paradoxically, under "lockdown", we now give up freedom, for which we sacrificed lives, to save them.

This is the first time that the entire Italian people place all their hopes for the future in the ability of our scientists, to prevent further contagion, find therapies, and finally eradicate COVID-19. All of our hearts and minds are with the "*angeli della scienza*".

As in times of war, the best scientific minds are at the service of the nation. All those with the competences and expertise that respond to the COVID-19 emergency have joined in the efforts. Leading scientists have put their other research on hold, to devote all their energies to combating COVID-19.

As scientometricians, recognizing our distance from the research fields at the front of the response, we have still asked ourselves how we might assist. This study presents our humble answer: developing a methodology to assess the scientific wealth of territories in the medical specialties most directly involved on the COVID-19 front, by measuring the intellectual or "knowledge" capital of territories and the relative performances of their academics. We apply the methodology to the case of Italy, which suffers from a chronic north-south economic divide, which reflects also in the health care infrastructure. We offer this rear-guard contribution in the hopes that one day someone could use the results to inform research policy decisions and actions that reduce the Italian disparities, or draw on the method for analyses in other countries.

The lethal force of COVID-19 has struck most heavily in northern Italy. The central area, and above all the south, were hit with much lesser intensity. As the initial dramatic effects were observed in the north, the government imposed strict restrictions that substantially shielded the south from the vigorous spread of the contagion. Given the north-south socio-economic divide so characteristic of Italy (Daniele & Malanima, 2011), commentators agree that if it were the south that had first been hit, the human catastrophe would have been far more devastating. This is because the north-south divide also extends to the areas of health infrastructure, in terms of its capacity, efficiency, effectiveness and quality of care (SVIMEZ, 2019).

The roots of the north-south divide are remote in time, deriving from many factors in history, economy, political geography, culture and society (Trigilia, 2012; Felice, 2014; Daniele & Malanima, 2014). Paradoxically, it is precisely the economic backwardness of the south that has saved it from the pandemic tsunami. The north, more densely populated, industrialized and dynamic, and therefore more inclined to international commerce, first imported the virus from Germany, through the international travel of a Lombard businessperson. The high employment rate and dense networks of interconnecting production chains, which are the sources of northern wealth, then offered the virus an ideal field for rapid contagion, before travel and other restrictions came into force. One of the specific aims of the restrictions was to block the north-south mobility of the virus, also explaining why inter-regional travel was still banned through “phase 2” of the national strategy (4 May 2020), when there were other relaxations of social constraints and permissions to resume work in a number of productive sectors.

[1] The manuscript was completed on 24 June 2020. At submission for publication, Italy ranked fourth for number of COVID-19 deaths, after the USA, UK, and Brazil.

The differing capacities of response to a pandemic such as COVID-19 depend on the state of the health care infrastructure, but also on the performances of the medical personnel, and therefore on the underlying accumulated scientific knowledge. In fact, the new scientific knowledge produced is transferred to practitioners through higher education and academic literature, and feeds continuous updating for those in active practice. Professors in medical schools differ from those of other disciplines, because in addition to doing research and teaching, they also carry out clinical activities. In this way they promote the immediate application of the most advanced knowledge and its rapid transfer among hospital colleagues. Last but not least, professors are called by the national and regional governments as members of committees to advise on the prevention and response to health care emergencies. Indeed, we have witnessed different strategies and initiatives across the Italian administrative regions in the specific case of COVID-19.

The periodic reports and other inquiries into the state of the national health care system typically lack statistical information on the scientific wealth in the medical disciplines. Instead they focus on tangible assets: the buildings, beds, equipment and the specialized personnel, and on their distribution across territories. In the case of COVID-19 pandemic, the debate on the differing regional capacities (e.g. numbers of personnel, PPEs, ventilation equipment) once again focuses on these tangible aspects.

Our current study aims to address this fault. With the help of specialists, we first identified the fields of medical sciences engaged on the front lines against the COVID-19 emergency. We then attempt to measure the relevant scientific wealth of each territory, by the “knowledge capital” (KC) produced, and the research performance of professors. In operational terms, we measure: i) the relevant total knowledge produced by each single territory in the latest measurable five-year period; but also ii) the competences of the professors in the identified fields, in terms of their research performance. The rationale underlying the latter measurement is that the university professors function as: direct contributors to the KC; health practitioners; the educators and advisors of other practitioners; consultants informing the health-sector actions of local governments. Their performance then, together with the available KC, would be play an important role in the effectiveness of the territories in the face of COVID-19.

It might be objected that the non-proprietary knowledge capital is a public good, and therefore easily transferable across territories. To a certain extent this is true, but a share of new knowledge produced is tacit, and can be applied only personally or transferred to others through personal interactions (Allen, 1984). Furthermore, the ease of transfer of the knowledge encoded in scientific publications is also affected by geographic distance. The existence of a geographic proximity effect has indeed been demonstrated, whether the transfer concerns knowledge embedded in patents (Jaffe, Trajtenberg, & Henderson, 1993) or in publications (Matthiessen Wichmann, Winkel Schwarz & Find, 2002; Börner, Penurnarthy, Meiss & Ke, 2006; Liu & Rousseau, 2010; Pan, Kaski & Fortunato, 2012). In our own research we have demonstrated that in Italy too, geographic proximity favors knowledge flows (Abramo, D’Angelo & Di Costa, 2020a). The geographic proximity effect has been shown to vary across research fields, including when controlling for the similarities of knowledge bases (cognitive proximity) in the territories involved in the transfers. In particular, in the medical and health sciences the average distance between domestic knowledge producers and users is 361.5 km, while the maximum is 1,119 km (Abramo, D’Angelo & Di Costa, 2020b). Although the geographic proximity effect decays over time (Abramo, D’Angelo & Di Costa, 2020c), it is particularly important when prompt response is needed, as in the COVID-19 emergency. The Italian case is also one of scarce domestic labor mobility, where students, academics, public and private sector workers seize on any opportunities close to their context of origin.

We are aware that the response to a pandemic must be a unitary effort of the overall national health system, not just of the specialists directly engaged at the front, and that success depends heavily on the intensity of collaboration across regional and national boundaries, and the sharing of knowledge of all kinds. However, our study could also be expanded to other medical research fields. The method of the study and the current results could already be useful to Italian policy makers at both the national and regional levels, when formulating plans to equitably respond to health care needs. The methods could also be replicated in other countries.

The next section of the paper profiles the Italian health care system, with particular emphasis on territorial inequities. Section 3 presents the methodology used to map relative territorial scientific strengths and weaknesses in the fields at the front of the COVID-19 emergency, and the data used. Section 4 presents the empirical evidence. In Section 5 we comment on the findings and draw conclusions.

## 2. Equity in the Italian health care system

The very recent Health Equity Status Report (WHO, 2019) provides an overall view of the evidence for “health inequities”, both between and within European countries, with the intention of informing government policy. In a subsequent editorial, The Lancet Public Health issued a call to action to reduce health inequalities (The Lancet Public Health, 2019). As yet there has been little response from Italian scholars (Paterlini, 2019; La Colla, 2019). Here we are limited to a brief profile of the Italian health care system, remarking territorial disparities when reported in the official statistics.

The Italian National Health Service is highly decentralized. The central government is responsible for formulating general guidelines and funding health budgets, for establishing the essential levels of care and monitoring their equal provision. Within the national framework, the 20 administrative regions have substantial autonomy in planning and organizing their health systems. This results in different organizational models and capacities, which in turn have given rise to different consequences at the onset of the COVID-19 emergency.

Among European nations, life expectancy in Italy is second to Spain, although unevenly distributed across regions and socio-economic groups (EU, 2019). At 24% age 65 and over (PRB, 2020), the share of older population is very large, contributing to the high death rates in cases of COVID-19 infection, although with variability among the regions. Another contributing factor has been the lack of availability of intensive care beds, exacerbated by continuous budget cuts since 2000, again with inequities across the regions.

Health spending per capita is about 15% below the EU average, and represents 8.8% of GDP, one point below the EU average. The number of doctors per capita is higher than EU average, but that of nurses is lower (EU, 2019). Health spending per capita is 25% higher in the north compared to the south, where on the contrary, demand for health services is higher. The availability of hospital beds per 100,000 residents is 791 in the center-north and 363 in the south (SVIMEZ, 2019).

Although the constitution indicates the right to equal access and quality of health services, regional differences persist (CERGAS-Bocconi, 2019). Health services are in principle universally available, largely free of charge, however the realities involve lack of local offer, wait times of varying duration and differing quality in the services, with all these conditions experienced more heavily in the south. The result is that southern patients increasingly seek services in the north, in the hope of obtaining quicker and better quality treatment.

Government is discussing a reform aimed at increasing regional tax autonomy, with the intent of increasing capabilities and public accountability in the regional administrations, and in these ways incentivizing performance (Bardhan, 2002; Weingast, 2009). The evidence suggests that such strategies would not affect inequities at the national level, but could help to reduce them within the individual regions (Di Novi, Piacenza, Robone, & Turati, 2019).

Very little is known about inter-regional disparities in medical sciences research. Several studies have investigated the north-south divide in the university system (Banfi & Viesti, 2015; Viesti, 2015; Abramo, D'Angelo, & Rosati, 2016; Viesti, 2016; Grisorio & Prota, 2020), however none of these focus on the medical sciences, and none deal with this area across the entire national research system. Our work should unveil any regional disparities in scientific wealth in the specialties directly concerned with the COVID-19 pandemic, and could be readily expanded to any other medical sciences areas.

## 3. Data and methods

We investigate the territorial disparities in KC and the performance of university professors relevant to the COVID-19 emergency, proceeding by four steps: first, identification of the medical specialties at the front of the emergency response; next, measurement of the KC in the specialties; third, measurement of the research performance of the university professors in the specialties; finally, delineation of territories and assignment of the KC and university professors to each of them, enabling the final regional comparisons.

### 3.1 Medical specialties relevant to COVID-19

The prevention, diagnosis, and treatment of COVID-19 pandemic requires an interdisciplinary, pan-systemic approach. However, some medical specialties are more directly implicated than others. To identify these, we first observed the specializations represented in the ad-hoc advisory committee established by the national government at the onset of the COVID-19 crisis. Additionally, we surveyed the opinions of 10 physicians with different specializations and occupations (e.g. universities, public and private hospitals, family practitioners). To do this we provided each respondent with two classification schemes of research specializations, asking them to identify those they considered on the "front line" of COVID-19 response, and those in the "second ranks".

One of the classification schemes provided is that of the Ministry of University and Research (MUR), used to officially categorize each professor employed in an Italian university by their research field. This scheme is composed of a total 370 "scientific disciplinary sectors" (SDSs), grouped into 14 "university disciplinary areas" (UDAs). The other scheme provided is the Web of Science (WoS) subject category (SC), consisting of 254 SCs, used to classify journals, and indirectly publications. We then reconciled the selections made by all respondents in the two different schemes.

We are aware that the perception of which medical specialties would be on the front lines of an emergency response would be variable, depending on the nature of the crisis, or varying with the evolution of a particular epidemic, or given a specific national context. The methodology we use thus allows adjustment of the selection of specialties, according to the case and context of interest.

### 3.2 Measurement of "knowledge capital"

Medical knowledge is produced by the different structures composing the Italian research system, most importantly universities, public research institutes, private companies and nationally designated research hospitals. To measure the KC produced by this infrastructure, we recur to its research output encoded in the publications indexed in the Italian National Citation Report (I-NCR) extracted from WoS by Clarivate Analytics.

KC depends both on the size of research (production factors) in the territory and on the ability of scientists to conduct research. The basic unit of analysis to measure KC is the publication.

The new knowledge encoded in publications is not all of equal value. Scientometricians measure the value of publications in terms of their scholarly impact, using citation-based indicators (Bornmann & Daniel, 2008; Tahamtan, Safipour Afshar, & Ahamdzadeh, 2016; Tahamtan & Bornmann, 2018; Abramo, 2018). Because citation behavior varies across research fields (Mingers, 2008; Baumgartner & Leydesdorff, 2014), field normalization is applied to compare the citation-based impact of publications of different fields. A vast literature on field normalization of citations proposes both parametric (Waltman & Van Eck, 2013) and non-parametric techniques (Mingers & Leydesdorff, 2015). Comparing a number of normalization approaches, Abramo, Cicero, and D'Angelo (2012) propose that the best normalization is obtained by dividing the number of citations of a publication by the average number of citations of all cited publications of the same year and field. We adopt this approach, assigning each publication to the WoS-designated SC of the hosting journal.

The life cycle of any scientific article begins at publication and ends on the date of its last citation. Any measurement of its impact before the life cycle is over is thus a prediction of the final impact: the longer the window of observing citations, the more accurate the prediction. To improve the power of predicting overall impact of publications, we adopt a weighted combination of citation counts and journal impact factor (IF), where the weights, per year of publication and SCs, are provided in Abramo, D'Angelo, and Felici (2019).

### 3.3 Measurement of professors' research performance

We measure professors' performance by their research productivity. Productivity, the quintessential indicator of efficiency in any production system, is commonly defined as the rate of output per unit of input. But because publications (output) have different values (impact), and the resources available for research are unequal across individuals and organizations, an appropriate definition of productivity in research systems is: the value of output per euro spent in research.

The FSS (fractional scientific strength) indicator is a proxy measure of research productivity. A thorough description of the FSS indicator and the underlying theory can be found in Abramo and D'Angelo (2014).[2]

To measure the yearly average research productivity of Italian academics, we use the following formula:[3]

$$FSS = \frac{1}{\left(\frac{w_r}{2}+k\right)} \cdot \frac{1}{t}\sum_{i=1}^{N} c_i f_i \qquad [1]$$

where:
$w_r$ = average yearly salary of professor[4]
$k$ = average yearly capital available for research to professor[5]
$t$ = number of years of work by the professor in period under observation
$N$ = number of publications by the professor in period under observation
$c_i$ = weighted combination of field-normalized citations and field-normalized impact factor associated to publication $i$[6]
$f_i$ = fractional contribution of professor to publication $i$.

The fractional contribution depends on the position of the authors in the publication byline and the character of the co-authorship (intra-mural or extra-mural). For publications resulting in intra-mural co-authorship 40% is attributed to first and last author, the remaining 20% is divided among all other authors. For publications resulting in extra-mural co-authorship, 30% is attributed to the first and last authors, 15% to the second and last authors but one, the remaining 10% is divided among all others.

The FSS score of professors belonging to different fields (SDSs, in the Italian university classification scheme) cannot be compared directly. In fact: i) scientists' intensity of publication remarkably varies across fields (Sandström & Sandström, 2009; Lillquist & Green, 2010; Piro, Aksnes & Rørstad, 2013; Sorzano, Vargas, Caffarena-Fernández, & Iriarte, 2014), and this particularly true in Italy (D'Angelo & Abramo, 2015); ii) the intensity of collaboration, i.e. the average number of co-authors per publication, also varies across fields (Glanzel & Schubert, 2004; Yoshikane & Kageura, 2004; Abramo, D'Angelo, & Murgia, 2013).[7]

To avoid distortions resulting from direct comparison of the performance of scientists belonging to different SDSs, and in comparisons at higher levels of aggregation (i.e. the territory), we normalize FSS scores to the average score of all professors of the same field, excluding those with nil score.[8] To exemplify, an FSS score of 1.10 means that the professor's performance is 10% above average, in their own SDS. In the following tables, figures and text, all FSS scores are normalized.

[2] This description of the measurement of research performance is similar to that in other publications by the authors.
[3] An underlying assumption is that labor and capital equally contribute to production.
[4] We halve the labor costs, assuming the allocation of 50 per cent of professors' time to non-research activities.
[5] Sources of input data and assumptions adopted in the measurement are found in Abramo, Aksnes, and D'Angelo (2020).
[6] Journal IF is measured at year of publication.
[7] The intensity of collaboration and impact of resulting publication are found to be positively correlated, at global level (Larivière, Gingras, Sugimoto, & Tsou, 2014) and in Italy (Abramo & D'Angelo, 2015), especially for international collaborations, at global level (Adams, 2013; Kumar, Rohani, & Ratnavelu, 2014) and in Italy (Abramo, D'Angelo, & Murgia, 2017).

Productivity (normalized FSS) is size independent and measures the professors' performance in conducting research. All others being equal, such performance should also reflect on the quality of education, of medical practice, and of consulting in emergency contexts.

Differently from the measurement of KC, the basic unit of analysis for the measure of research productivity is the scientist. Because of this, we need to link each professor with their publications in the period under observation. However the raw data for institutional and author identification available in the WoS are open to ambiguities, for example due to different naming of institutions, uses of full names or initials, and homonyms of first and last names. For this, we have developed a complex algorithm for application to WoS data, able to reconcile the author's institutional affiliation and disambiguate their true identity.[9] The disambiguation algorithm can only be used with university professors, since Italy does not maintain databases of other national research personnel. For this reason we cannot extend the measurement of performance to other medical scientists who may be operating in the regional systems.

### 3.4 Allocating knowledge capital and professors' performance to territories

Under the EU statistical classification, Italy is divided into: 5 macro-regions (northwest, northeast, center, south, and islands (NUTS 1 level); 19 regions and two autonomous provinces, composing the 20th region (NUTS 2), and 110 provinces (NUTS 3).[10] The provinces are further divided into local administrative unit (LAUs, or municipalities), of which Italy counts 11,107.[11] Municipalities, provinces and regions have elected governments, with authority, responsibility and budgets in the area of health services. Macro-regions are a statistical concept, without legal or operative status.

The northwest macro-region is comprised of the regions of Valle d'Aosta, Piedmont, Liguria, and Lombardy; the northeast of Veneto, Trentino Alto-Adige, Friuli Venezia Giulia, and Emilia Romagna; the center of Tuscany, Marche, Umbria, and Lazio; the south comprises Abruzzo, Molise, Campania, Basilicata, Puglia, and Calabria; the island macro-region includes Sardinia and Sicily.

The KC is assigned to the territories of production as follows.

Taking the I-NCR, which registers all publications with "Italy" in the affiliation list, for each publication, we reduce all addresses to city + country (e.g. "Rome, Italy"). Each I-NCR "city" is then assigned to its LAU, according to the official lists of the National Institute of Statistics (ISTAT).[12]

---

[8] Abramo, Cicero, and D'Angelo (2012) demonstrate that the average of the distribution (excluding nil values) is the best-performing scaling factor. In comparing Italian professors, the "average" used to rescale original distributions is calculated on Italian performance distributions.

[9] The harmonic average of precision and recall (F-measure) of authorships, as disambiguated by the algorithm, is around 97% (2% margin of error, 98% confidence interval).

[10] https://eur-lex.europa.eu/legal-content/EN/ALL/?uri=CELEX:02003R1059-20180118&qid=1519136585935, last accessed on 24 June 2020.

[11] https://ec.europa.eu/eurostat/web/nuts/local-administrative-units, last accessed on 24 June 2020.

[12] https://www.istat.it/it/archivio/6789, last accessed on 24 June 2020.

Most publications are coauthored by scientists from different territories (LAUs), and for this there could be different approaches in allocating the publications:

i) to each of the territories of the institutions in the address list;

ii) to a single territory, e.g. by frequency of authors (or institutions) of the territory in the address list, by affiliation of the corresponding author, by affiliation of the first or last author in non-alphabetically ordered bylines;

iii) by fractionalizing the publication by the number of territories, institutions or authors.

We adopt the fractional counting method, assigning the publication to each territory in proportion to the weighted number of authors belonging to a specific territory. As said above, the weights depend on the positions of the authors in the byline and the character of the co-authorship (intra-mural or extra-mural).

To account for cases of single authors with multiple affiliations, we again adopt a fractional counting method. For an author with *m* affiliations, we assign 1/*m* to each of their bibliometric addresses (LAUs).

As an example, the box below shows the byline of the publication with DOI 10.1016/j.atherosclerosis.2013.10.017.

Scoditti, E[1], Nestola, A[1], Massaro, M[1], Calabriso, N[1], Storelli, C[2], De Caterina, R[3,4], Carluccio, MA[1]
[1] National Research Council (CNR), Institute of Clinical Physiology (IFC), Lecce, Italy
[2] Department of Biological and Environmental Science and Technologies, University of Salento, Lecce, Italy
[3] "G. Monasterio" Foundation for Clinical Research, Pisa, Italy
[4] "G. d'Annunzio" University and Center of Excellence on Aging, Chieti, Italy

The extramural character of the collaboration between the authors is evident. The "Lecce" LAU accounts for all the authors but "De Caterina, R", the last author but one, who shows 2 distinct affiliations ("G. Monasterio" in Pisa and "G. d'Annunzio" in Chieti). In this case, applying all the above rules, we assign a fraction of 0.075 (0.15x0.5) to both Chieti and Pisa, and 0.850 to Lecce.

The KC of a territory *t* then is the sum of the fractional scholarly impact of publications authored by scientists located in that territory:

$$KC = \sum_{i=1}^{N} c_i \sum_{j=1}^{a_i} f_j \qquad [2]$$

where:

$f_j$ = fractional contribution of the j-th author

$a_i$ = number of authors of publication *i*, affiliated to organizations located in the territory *t*

$c_i$ = weighted combination of field-normalized citations and field-normalized impact factor associated to publication *i*

*N* = number of publications authored by at least one author affiliated to organizations located in the territory *t*

Because territories can differ greatly in population size, we will also measure the knowledge capital per (million) capita, $KC_{PC}$.

Similarly to the measurement of professors' performance, to avoid distortions when aggregating $KC_{PC}$ of different specialties at territorial level, we first normalize $KC_{PC}$ in each specialty to the average national $KC_{PC}$. To exemplify, a $KC_{PC}$ score of 1.10 in a specialty means that in that territory the $KC_{PC}$ is 10% above national average.

In the following tables, figures and text, all $KC_{PC}$ scores are normalized.

With the disambiguation of the professors' names and university affiliations (section 3.3), their research performance is readily allocated to the LAUs where universities are located, and can be aggregated at the higher NUTS levels.

### 3.5 Data

We apply several criteria in choosing the period of observation, or "citation window". The interval cannot be too long, because of the rapid obsolescence of knowledge in the medical sciences, yet not so short as to result in biases from random fluctuations in values or data collection. The closing date must be sufficiently recent for observation of the current scientific wealth, yet not so recent as to jeopardize robust measurement of its value. A good compromise is a five year period, from 2014 to 2018. We observe and count the citations in this window as of 31/12/2019.

We extract the data on the professors of each Italian university from the database on personnel maintained by the MUR, including their first and last name, gender, university affiliation, field classification and academic rank, at the close of each year.[13] The MUR recognizes 98 Italian universities with authority to issue degrees. Of these, 68 - distributed across all the regions but Valle D'Aosta (population 126,000) - employ 5614 professors in one or more of the 20 medical specialties under observation.

Table 1 shows the knowledge production in the specialties, as encoded in 96034 publications indexed in the I-NCR, and its territorial distribution. Specialty 2, Biochemistry & molecular biology, shows the highest number of publications; those in General internal medicine received the highest number of citations; Clinical neurology shows the highest number of authorships. Sixteen specialties are present in all regions, but publication in three specialties (Anesthesiology and emergency medicine, Pharmaceutical biology, Medical laboratory technology) is absent in Valle d'Aosta,[14] Nursing is absent in Molise. The situation is less uniform at the level of the provinces. Around one third lack KC in Pharmaceutical biology and in Nursing; in contrast, KC is present in 107 of the 110 provinces for Pharmacology & pharmacy, Public health and General internal medicine.

---

[13] http://cercauniversita.cineca.it/php5/docenti/cerca.php, last accessed on 24 June 2020.

[14] As noted, the universities of Val d'Aosta have no professors in the front-line specialties, however in 17 of the fields, articles have been published by authors from this region. These could be by professors classified in other specialties, or more likely by researchers employed in organizations other than universities.

*Table 1: Bibliographic data by specialty and territory, I-NCR data 2014-2018*

| Specialty* | Publications | Citations | Authorships | Provinces | Regions |
|---|---|---|---|---|---|
| 1 | 1806 | 17960 | 9642 | 94 | 19 |
| 2 | 12119 | 181138 | 65368 | 101 | 20 |
| 3 | 411 | 3563 | 2059 | 75 | 19 |
| 4 | 8163 | 159801 | 54850 | 101 | 20 |
| 5 | 4426 | 54833 | 26063 | 89 | 20 |
| 6 | 6649 | 74129 | 39875 | 101 | 20 |
| 7 | 11476 | 130317 | 64702 | 107 | 20 |
| 8 | 4684 | 41175 | 23376 | 106 | 20 |
| 9 | 10371 | 168090 | 60891 | 107 | 20 |
| 10 | 3732 | 52724 | 19610 | 98 | 20 |
| 11 | 3863 | 47363 | 24090 | 104 | 20 |
| 12 | 7166 | 235250 | 41420 | 107 | 20 |
| 13 | 1345 | 14577 | 8067 | 87 | 20 |
| 14 | 9991 | 128211 | 55636 | 104 | 20 |
| 15 | 11075 | 140867 | 67091 | 106 | 20 |
| 16 | 7235 | 106339 | 44654 | 105 | 20 |
| 17 | 2407 | 24591 | 14548 | 98 | 20 |
| 18 | 4221 | 32687 | 23982 | 99 | 20 |
| 19 | 645 | 3924 | 3155 | 74 | 19 |
| 20 | 1080 | 10983 | 5864 | 80 | 19 |
| *Total* | *96034* | *1403167* | *654943* | *110* | *20* |

* 1, Anesthesiology and emergency medicine; 2, Biochemistry & molecular biology; 3, Pharmaceutical biology; 4, Cell biology; 5, Medicinal chemistry; 6, Radiology, nuclear medicine & medical imaging; 7, Pharmacology & pharmacy; 8, Public health; 9, Cardiac & cardiovascular systems; 10, Respiratory system; 11, Infectious diseases; 12, General internal medicine; 13, Virology; 14, Microbiology; 15, Clinical neurology; 16, Immunology; 17, Pathology; 18, Pediatrics; 19, Nursing; 20, Medical laboratory technology.

## 4. Results

Before reporting the current analyses, we show the findings of a previous work identifying the strengths and weaknesses of Italian academic research at field level (Abramo & D'Angelo, 2020). We used two indicators, i) the average performance of professors in each SDS in producing articles that achieve high citation at the global level, and ii) the share of such professors out of total in the SDS. In each of the 218 SDSs where bibliometrics can be reasonably applied (out of 370 total), we measured the average research performance by two main indicators: the ratio of i) highly cited articles, and ii) numerosity of top scientists relative to research expenditures in the SDS. We define highly cited publications as those that place in the top 5% or 10% of the world citation rankings for those indexed in the WoS, of the same year and subject category. Top scientists are the professors with a total fractional counting of highly cited articles that exceeds a chosen threshold. Each SDS was then ranked by the average of their ranks for the two indicators. Table 1 (last column) presents the percentile performance rank for the 20 medical specialties currently under observation, out of the total 218 SDSs analyzed for the Italian context. Within the 20 front-line specialties, we find 3 in the top decile among Italian university SDSs, 10 in the top quartile, 16 above average, and none in the bottom quartile.

***Table 2: Specialties on the front lines against COVID-19: relative size and professors' research performance (Pctl 100 = top) vis-à-vis all bibliometric fields (218) in the Italian academic system (percentage share of total staff in the dataset, in brackets)***

| Specialty | No. of Professors* | Size (Pctl) | Research performance (Pctl) |
|---|---|---|---|
| Pathology | 489 (8.7%) | 93.5 | 99.1 |
| Cardiac & cardiovascular systems | 247 (4.4%) | 79.1 | 95.0 |
| Cell biology | 244 (4.3%) | 78.7 | 92.7 |
| Respiratory system | 105 (1.9%) | 42.5 | 89.9 |
| Microbiology | 120 (2.1%) | 49.5 | 89.0 |
| Biochemistry & molecular biology | 154 (2.7%) | 61.1 | 87.1 |
| Clinical neurology | 350 (6.2%) | 88.4 | 84.4 |
| Public health | 429 (7.6%) | 93.0 | 82.4 |
| General internal medicine | 739 (13.2%) | 97.6 | 79.3 |
| Pharmacology & pharmacy | 637 (11.3%) | 95.8 | 77.9 |
| Medicinal chemistry | 428 (7.6%) | 92.5 | 70.6 |
| Anesthesiology and emergency medicine | 216 (3.8%) | 76.8 | 70.1 |
| Medical laboratory technology | 141 (2.5%) | 56.9 | 65.9 |
| Virology | 310 (5.5%) | 85.6 | 64.1 |
| Pediatrics | 332 (5.9%) | 87.5 | 63.6 |
| Infectious diseases | 139 (2.5%) | 56.0 | 49.4 |
| Radiology, nuclear medicine & medical imaging | 320 (5.7%) | 86.1 | 48.4 |
| Immunology | 110 (2.0%) | 45.8 | 42.9 |
| Nursing | 32 (0.6%) | 7.8 | 27.7 |
| Pharmaceutical biology | 72 (1.3%) | 21.7 | 25.4 |
| Total | 5614 | | |

** On staff for at least three years in the 2014-2018 period.*

Turning to the current study, in the subsequent sub-sections we first map the knowledge capital in each specialty and territory. Next we report the analysis of the average performance of professors in their respective research specialties, for each territory. Finally, we integrate these two analyses to obtain an overall view of the positioning of all the territories within the national context, in terms of the two analyses: average professor's performance and knowledge capital per capita.

## 4.1 Knowledge capital by specialty in each territory

As noted, the COVID-19 contagion spread through Italy in a remarkably uneven manner. As soon as the first outbreaks were observed, in the north, the national government imposed containment measures on individual communities. Very soon after the entire country was placed on "lockdown", banning any movement except for essential reasons. This meant that the territories in central and especially southern Italy were largely protected from the pandemic, as demonstrated in Figure 1. The figure shows the distribution of COVID-19 cases by province (NUTS 3), per thousand residents, as of 18 May 2020. At that date, 225,886 cases had already been recorded in Italy (3.7 per 1,000 residents).

Four of the five leading provinces for incidence of contagion are in Lombardy (Cremona, 17.6; Lodi, 14.6; Brescia, 11.2; Bergamo, 11.2); the fifth is in Emilia Romagna (Piacenza, 15.4). Only nine of the total 47 northern provinces had incidences lower than national average (all these in the northeast).

In the center, only three provinces had incidences higher than the national average (two in the region of Marche, one in Tuscany); in the south only one province was above national average (Pescara, in Abruzzo). The five provinces with the lowest incidence (less than 10% of the national average) are all on the islands: three in Sicily and two in Sardinia.

***Figure 1: Provincial distribution of COVID-19: infected residents per thousand residents (data as of 18 May 2020)***

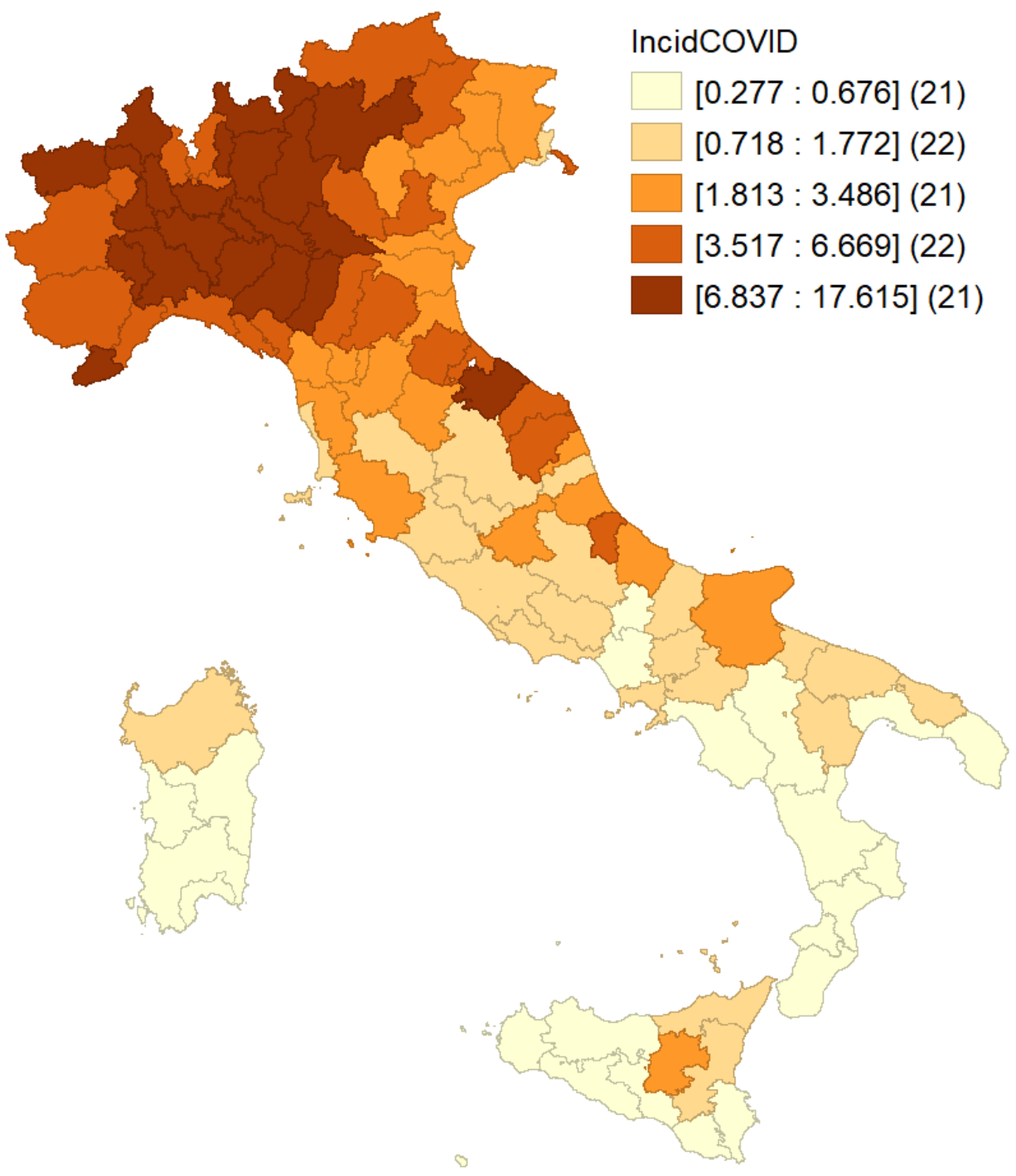

By preparing similar maps of the provincial distribution of $KC_{PC}$, FSS or both, we can contrast the impact of the coronavirus and the scientific wealth in support of the response. Figure 2 shows, for example, the provincial distribution of $KC_{PC}$ for the "Public health" specialty, in particular. Among the 18 provinces with scores of $KC_{PC}$ near zero (less than 3% of the national average), eight are in the south and four in the center. The remaining six are in the north, more precisely: two in Lombardy, two in Piedmont and two in Veneto.

Eight of the 110 Italian provinces register a $KC_{PC}$ in Public health at least double the national average: one province each in the regions of Friuli Venezia Giulia, Emilia Romagna, Lombardy and Lazio and three in the region of Tuscany. These data reveal a significant geographical concentration of research activity in Public health. Four of the eight provinces concerned have a resident population of more than 1 million units (Bologna, Rome, Milan, Florence); the remaining four host important national research centers in the medical sciences (Trieste, Pisa, Siena, Padua).

***Figure 2: Provincial distribution of knowledge capital per capita ($KC_{PC}$) for the "Public health" specialty***

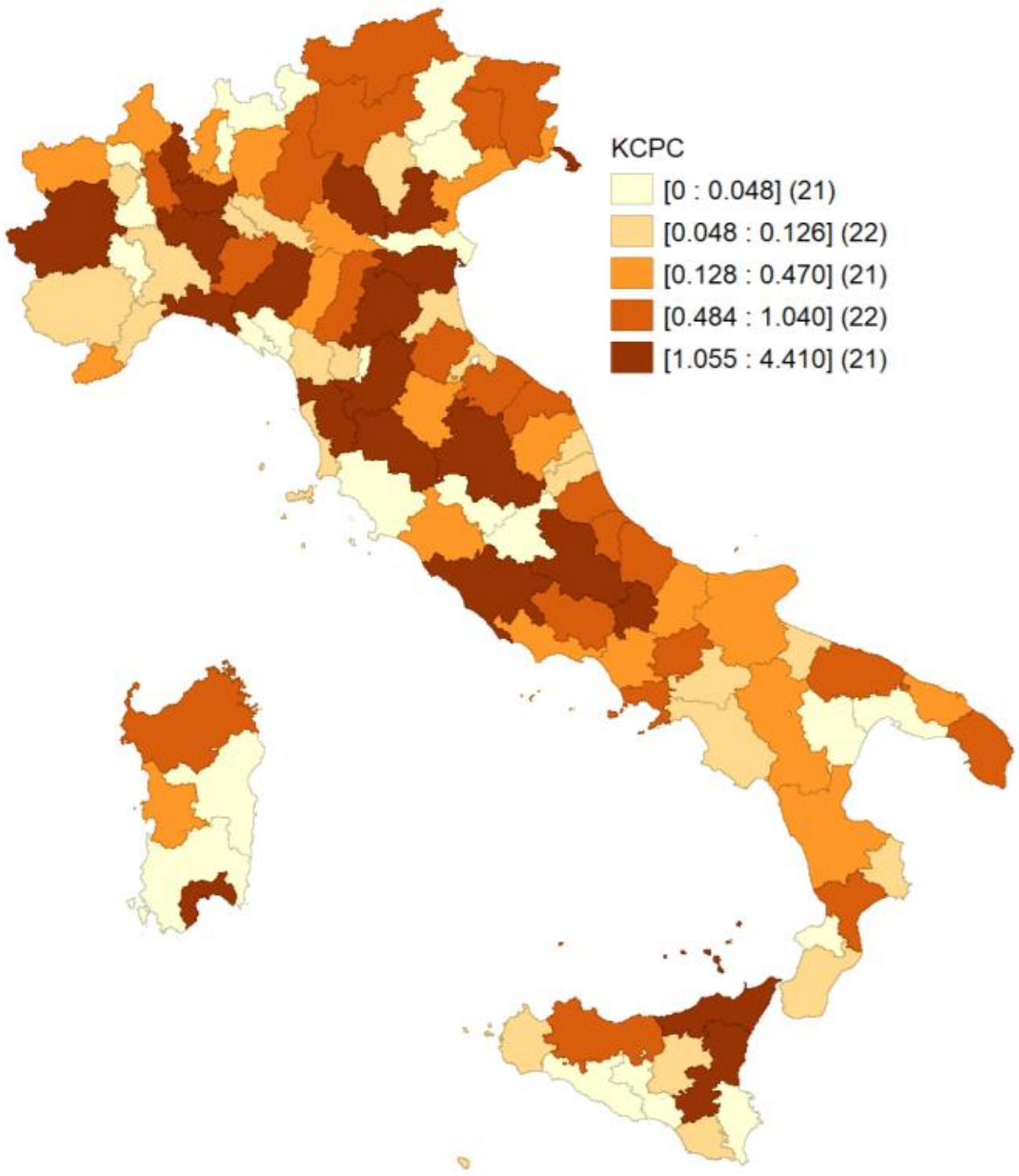

In the expectation of providing useful regional and national policy indications, we aggregate the distribution of $KC_{PC}$ at the regional level. In Table 3, the last column shows the simple average of the scores for all 20 COVID-19 front-line specialties, including the instances of $KC_{PC} = 0$, i.e. nil KC in a region.[15]

The regions of the south and islands all achieve $KC_{PC}$ scores in the 20 specialties that are lower than national average, with the exception of Abruzzo (1.05). The score for Basilicata is particularly low: this is a small region (population 560,000), but average $KC_{PC}$ (0.19) is still less than a fifth of the national average.

Valle d'Aosta (population 120,000, in the Italian northwest) records an even lower score (0.11). Other northern regions with $KC_{PC}$ lower than the national average are Trentino Alto Adige (0.59) and Piedmont (0.72). The remaining five regions of northern Italy show scores above average, with Friuli Venezia Giulia the highest (1.27). However, it is Tuscany and Lazio (two central regions) that record the national absolute maximums of $KC_{PC}$ (respectively 1.45, 1.71), reflecting the concentrations of public and industrial medical research centers in these two regions.

Reading across the rows, we can observe the regional distribution of $KC_{PC}$ by specialty: in Anesthesiology and emergency medicine, there is no record of research activity in Valle D'Aosta, over the period under observation, and almost none in Basilicata ($KC_{PC} = 0.02$). The maximum figures for this specialty are in Lombardy (1.65), then Liguria and Marche: all three are regions where COVID-19 hit particularly hard. In Public health, the most important concentrations of activity are in Lazio ($KC_{PC} = 2.31$) and Tuscany (1.48). Lombardy, Emilia Romagna and Liguria also recorded scores above the national average (1.24, 1.16 and 1.02). The regions at the bottom of the $KC_{PC}$ ranking in this specialty are Basilicata, Calabria and Puglia (all in the south), at $KC_{PC}$ respectively 0.10, 0.30, 0.44, and Valle d'Aosta (northwest) at 0.30.

[15] The method can be readily adapted to different selections of specialties, yielding the relevant average $KC_{PC}$ scores.

Basilicata, Calabria and Valle d'Aosta are unique in lacking any specialty with $KC_{PC}$ above the national average. The top performing region of the south is Abruzzo, with $KC_{PC}$ above the national average in 35% of the front-line specialties (7 of 20). In contrast, Emilia Romagna, in the north, exceeds the national average in all specialties but Immunology (0.89) and Nursing (0.90). Lazio and Tuscany, both in central Italy, show similar strong performance: above the national average in 18 (Lazio) or 17 (Tuscany) specialties, and for Tuscany, with absolute national records in Biochemistry & molecular biology (1.76), Pharmaceutical biology (2.73) and Medicinal chemistry (3.22).

***Table 3: Distribution of normalized knowledge capital per capita ($KC_{PC}$) in each specialty and region (data 2014-2018; 1.10 means 10% above average)***

| Region | Anesthesiology and emergency medicine | Biochemistry & molecular biology | Pharmaceutical biology | Cell biology | Medicinal chemistry | Radiology, nuclear medicine & medical imaging | Pharmacology & pharmacy | Public health | Cardiac & cardiovascular systems | Respiratory system | Infectious diseases | General internal medicine | Virology | Microbiology | Clinical neurology | Immunology | Pathology | Pediatrics | Nursing | Medical laboratory technology | Total |
|---|---|---|---|---|---|---|---|---|---|---|---|---|---|---|---|---|---|---|---|---|---|
| Abruzzo | 1.19 | 0.83 | 1.81 | 0.89 | 1.27 | 0.97 | 1.27 | 0.77 | 0.72 | 0.58 | 0.88 | 0.98 | 1.15 | 0.74 | 1.10 | 0.87 | 0.69 | 0.77 | 3.46 | 0.04 | 1.05 |
| Basilicata | 0.02 | 0.56 | 0.57 | 0.28 | 0.55 | 0.15 | 0.23 | 0.10 | 0.03 | 0.15 | 0.04 | 0.08 | 0.03 | 0.45 | 0.06 | 0.10 | 0.33 | 0.05 | 0.01 | 0.05 | 0.19 |
| Calabria | 0.26 | 0.59 | 0.78 | 0.71 | 0.80 | 0.19 | 0.69 | 0.30 | 0.44 | 0.11 | 0.23 | 0.57 | 0.04 | 0.39 | 0.42 | 0.30 | 0.42 | 0.11 | 0.02 | 0.23 | 0.38 |
| Campania | 0.35 | 1.15 | 0.83 | 1.06 | 1.41 | 0.71 | 1.01 | 0.59 | 0.82 | 0.63 | 0.42 | 0.69 | 0.50 | 1.01 | 0.71 | 0.70 | 0.78 | 0.79 | 0.17 | 0.38 | 0.74 |
| Emilia Romagna | 1.46 | 1.16 | 1.48 | 1.28 | 1.34 | 1.13 | 1.31 | 1.16 | 1.17 | 1.66 | 1.11 | 1.36 | 1.34 | 1.53 | 1.05 | 0.89 | 1.48 | 1.07 | 0.90 | 1.37 | 1.26 |
| Friuli Venezia Giulia | 1.11 | 1.41 | 0.17 | 1.13 | 0.91 | 1.22 | 0.95 | 0.96 | 1.11 | 0.81 | 1.42 | 0.91 | 2.28 | 1.03 | 0.42 | 1.08 | 0.80 | 1.64 | 5.04 | 0.93 | 1.27 |
| Lazio | 1.37 | 1.58 | 0.90 | 1.77 | 0.93 | 1.73 | 1.69 | 2.31 | 1.49 | 1.61 | 2.52 | 1.61 | 2.15 | 1.60 | 1.90 | 1.79 | 1.41 | 2.12 | 2.48 | 1.28 | 1.71 |
| Liguria | 1.56 | 0.80 | 0.53 | 1.03 | 1.12 | 1.22 | 1.15 | 1.02 | 0.49 | 1.12 | 1.50 | 1.06 | 0.57 | 0.86 | 1.19 | 2.55 | 0.75 | 1.82 | 2.35 | 0.64 | 1.17 |
| Lombardy | 1.65 | 0.89 | 0.92 | 1.09 | 0.58 | 1.57 | 0.97 | 1.24 | 1.47 | 1.61 | 1.15 | 1.47 | 1.11 | 0.83 | 1.58 | 1.45 | 1.15 | 1.18 | 0.81 | 1.40 | 1.21 |
| Marche | 1.55 | 0.85 | 0.91 | 0.81 | 0.78 | 0.16 | 0.85 | 0.56 | 0.37 | 0.43 | 0.74 | 0.74 | 0.13 | 0.91 | 0.64 | 0.33 | 0.93 | 0.55 | 0.15 | 0.30 | 0.63 |
| Molise | 0.28 | 0.74 | 0.76 | 0.83 | 0.55 | 1.20 | 1.02 | 0.55 | 2.02 | 0.65 | 0.19 | 0.89 | 0.25 | 0.78 | 1.77 | 0.74 | 0.53 | 0.34 | 0.00 | 0.02 | 0.71 |
| Piedmont | 0.81 | 0.66 | 0.60 | 0.77 | 0.43 | 0.77 | 0.67 | 0.78 | 0.73 | 0.83 | 0.59 | 0.77 | 1.46 | 0.76 | 0.51 | 0.45 | 0.91 | 0.69 | 0.82 | 0.39 | 0.72 |
| Puglia | 0.62 | 0.60 | 0.03 | 0.50 | 0.35 | 0.33 | 0.54 | 0.44 | 0.43 | 0.39 | 0.51 | 0.54 | 0.77 | 0.99 | 0.39 | 0.54 | 0.39 | 0.50 | 0.10 | 0.25 | 0.46 |
| Sardinia | 0.16 | 0.63 | 2.26 | 0.18 | 1.13 | 0.40 | 0.99 | 0.59 | 0.22 | 0.41 | 0.77 | 0.46 | 0.97 | 0.69 | 0.71 | 0.35 | 0.28 | 0.49 | 1.81 | 0.65 | 0.71 |
| Sicily | 0.56 | 0.66 | 1.05 | 0.69 | 1.23 | 0.51 | 0.96 | 0.72 | 0.65 | 0.49 | 0.41 | 0.67 | 0.35 | 0.67 | 0.65 | 0.76 | 0.62 | 0.72 | 0.25 | 0.40 | 0.65 |
| Tuscany | 1.11 | 1.76 | 2.73 | 0.98 | 3.22 | 1.35 | 1.60 | 1.48 | 1.71 | 1.27 | 1.38 | 1.09 | 0.73 | 1.42 | 1.26 | 1.39 | 1.27 | 1.28 | 0.87 | 1.18 | 1.45 |
| Trentino Alto Adige | 0.71 | 0.68 | 0.04 | 0.65 | 0.11 | 1.51 | 0.14 | 0.72 | 0.25 | 0.06 | 0.63 | 0.23 | 1.60 | 1.76 | 0.52 | 0.14 | 0.52 | 0.23 | 0.60 | 0.62 | 0.59 |
| Umbria | 0.30 | 1.18 | 0.89 | 0.88 | 1.50 | 0.50 | 0.96 | 0.99 | 0.80 | 1.20 | 1.07 | 1.31 | 1.25 | 1.16 | 0.87 | 1.15 | 0.93 | 0.65 | 1.15 | 0.21 | 0.95 |
| Valle d’Aosta | 0.00 | 0.05 | 0.00 | 0.04 | 0.01 | 0.07 | 0.00 | 0.30 | 0.02 | 0.44 | 0.01 | 0.01 | 0.09 | 0.04 | 0.06 | 0.00 | 0.62 | 0.07 | 0.02 | 0.00 | 0.09 |
| Veneto | 0.93 | 1.07 | 0.96 | 1.15 | 0.48 | 0.93 | 0.58 | 0.69 | 1.20 | 0.97 | 1.03 | 0.89 | 0.78 | 0.76 | 0.82 | 0.92 | 1.71 | 0.96 | 1.02 | 3.32 | 1.06 |

## 4.2 Average research performance of professors by specialty, in each territory

Aggregating the FSS data for all professors, by specialty and region, we obtain the scores for average performance shown in Table 4. The shaded scores indicate the specialties in which a region has less than five professors. The last column shows the region's average performance in all the specialties, and reveals the supremacy of Trentino Alto Adige: the professors of this region achieve an average performance 2.59 times higher than the national value, although there are only 14 professors, and these are active in only 5 of the 20 front-line specialties. Veneto, Lombardy and Emilia Romagna, three of the four regions most heavily impacted by the COVID-19 pandemic, also record research performance above the national average (respectively 1.27, 1.16, 1.09). Piedmont, the second region for number of infections, is slightly below the national average (0.99). Another northern region with performance below national average is Friuli Venezia Giulia (0.72).

On the other hand, the southern regions are all at the bottom of this ranking list, invariably showing FSS scores lower than the national average, with Molise at the bottom (0.57).

Besides the overall performance, by reading across the rows of Table 4 we can see the performance of each region in the different specialties – where it excels or lags behind the national average. Lombardy (northwest), for example, shows a performance at least 35% higher than the average in Virology (1.55), Respiratory system (1.35), Clinical neurology (1.40) and General internal medicine (1.38). In contrast, the same region shows performances significantly lower than the national average in Nursing (0.27), Pharmaceutical biology (0.43), Immunology (0.52), Medicinal chemistry (0.63) and Microbiology (0.71). Sicily (islands), instead, excels in Public health (1.65) and is second to Calabria (1.48) in Pharmacology & pharmacy (1.32), and to Veneto (2.02) in Medical laboratory technology (1.95).

Therefore, although numerically more frequent in the north, there are also excellent performances in the south, and vice versa for weak performances. In terms of our method, Table 4 certainly supports the identification of strengths and weaknesses in the specialties of the regional academic systems, necessary for the fight against the current pandemic.

***Table 4: Average research performance (normalized FSS) of Italian professors by specialty and region (1.10 means 10% above average)***

| Region | Macro region* | Population (x 1,000,000) | Total professors | Specialties covered (%) | Anesthesiology and emergency medicine | Biochemistry & molecular biology | Pharmaceutical biology | Cell biology | Medicinal chemistry | Radiology, nuclear medicine & medical imaging | Pharmacology & pharmacy | Public health | Cardiac & cardiovascular systems | Respiratory system | Infectious diseases | General internal medicine | Virology | Microbiology | Clinical neurology | Immunology | Pathology | Pediatrics | Nursing | Medical laboratory technology | Total |
|---|---|---|---|---|---|---|---|---|---|---|---|---|---|---|---|---|---|---|---|---|---|---|---|---|---|
| Abruzzo | S | 1.3 | 159 | 95 | 0.82 | 0.36 | 2.17 | 1.02 | 1.76 | 1.46 | 0.78 | 0.69 | 0.36 | | 0.29 | 0.93 | 1.15 | 0.09 | 1.04 | 0.39 | 0.45 | 2.44 | 1.08 | 0.91 | 0.91 |
| Basilicata | S | 0.6 | 8 | 30 | | 0.32 | 2.21 | 1.32 | 0.55 | | 0.34 | | | | | | | | | | | | | 1.87 | 0.93 |
| Calabria | S | 1.9 | 102 | 95 | 0.10 | 0.20 | 1.68 | 0.63 | 1.13 | 0.18 | 1.48 | 0.98 | 1.57 | 1.47 | 0.72 | 0.77 | 0.29 | | 1.27 | 1.12 | 0.76 | 0.53 | 0.06 | 0.95 | 0.97 |
| Campania | S | 5.8 | 512 | 100 | 0.34 | 0.51 | 1.38 | 0.61 | 0.77 | 0.70 | 1.05 | 0.90 | 0.92 | 0.42 | 1.32 | 0.90 | 0.84 | 0.35 | 0.85 | 0.71 | 0.77 | 1.05 | 0.31 | 0.60 | 0.83 |
| Emilia Romagna | NE | 4.5 | 547 | 100 | 0.99 | 0.62 | 0.49 | 0.91 | 0.92 | 1.66 | 0.93 | 0.99 | 1.46 | 0.99 | 1.70 | 1.04 | 1.00 | 1.71 | 0.97 | 1.88 | 1.75 | 0.48 | 0.49 | 1.21 | 1.09 |
| Friuli Venezia Giulia | NE | 1.2 | 102 | 85 | 0.60 | 1.38 | | 0.84 | 0.22 | 1.09 | 0.63 | 0.62 | 0.97 | | 0.67 | 0.73 | 0.57 | | 0.56 | 0.96 | 0.47 | 0.85 | 5.31 | 0.30 | 0.72 |
| Lazio | C | 5.9 | 824 | 100 | 0.31 | 0.50 | 1.06 | 1.21 | 0.95 | 0.74 | 1.08 | 0.60 | 0.44 | 1.27 | 0.55 | 0.71 | 0.72 | 0.62 | 0.90 | 0.72 | 0.61 | 0.70 | 1.32 | 1.03 | 0.74 |
| Liguria | NW | 1.6 | 151 | 100 | 5.77 | 0.32 | 0.79 | 0.76 | 0.60 | 1.18 | 0.74 | 1.31 | 0.76 | 1.13 | 2.07 | 0.67 | 1.48 | 1.38 | 0.86 | 0.51 | 0.83 | 1.30 | 2.08 | 0.46 | 1.03 |
| Lombardy | NW | 10.1 | 977 | 100 | 1.24 | 1.05 | 0.43 | 0.95 | 0.63 | 1.00 | 1.05 | 1.28 | 1.26 | 1.35 | 0.90 | 1.38 | 1.55 | 0.71 | 1.40 | 0.52 | 1.24 | 1.11 | 0.27 | 1.05 | 1.16 |
| Marche | C | 1.5 | 122 | 90 | 0.77 | 0.85 | 3.08 | 0.18 | 0.55 | 0.53 | 0.69 | 0.55 | 1.02 | 0.60 | 0.56 | 0.73 | 0.88 | 0.35 | 1.89 | | 1.20 | 3.66 | | 0.80 | 0.84 |
| Molise | S | 0.3 | 15 | 50 | | | | | | 1.93 | 0.82 | 0.38 | 0.46 | | | 1.11 | 0.47 | 0.27 | 0.69 | | 0.35 | | | 0.28 | 0.57 |
| Piedmont | NW | 4.4 | 272 | 100 | 1.34 | 1.49 | 1.02 | 1.15 | 0.56 | 1.15 | 0.63 | 1.13 | 1.59 | 1.35 | 2.48 | 0.91 | 1.02 | 1.16 | 0.98 | 0.40 | 1.12 | 0.64 | 0.65 | 1.22 | 0.99 |
| Puglia | S | 4.0 | 231 | 90 | 0.79 | 0.49 | | 0.94 | 0.70 | 1.20 | 0.78 | 0.72 | 1.20 | 0.78 | 0.44 | 0.92 | 0.35 | 0.53 | 1.50 | 0.12 | 0.46 | 1.08 | | 0.35 | 0.78 |
| Sardinia | I | 1.6 | 208 | 90 | 0.82 | 0.87 | 0.79 | 0.19 | 0.48 | 0.75 | 0.77 | 0.59 | 0.41 | 0.79 | 0.41 | 0.20 | 0.84 | 1.10 | 0.72 | 6.95 | 0.65 | 0.57 | | | 0.69 |
| Sicily | I | 5.0 | 469 | 100 | 0.50 | 1.07 | 0.30 | 0.74 | 0.79 | 0.65 | 1.32 | 1.65 | 1.02 | 1.15 | 0.57 | 0.81 | 0.43 | 1.01 | 0.46 | 0.82 | 0.95 | 1.02 | 0.38 | 1.95 | 0.88 |
| Tuscany | C | 3.7 | 454 | 100 | 0.80 | 1.08 | 0.75 | 0.71 | 2.45 | 1.18 | 1.04 | 0.85 | 1.32 | 0.44 | 1.51 | 0.85 | 1.91 | 0.95 | 0.80 | 0.85 | 0.42 | 1.07 | 0.66 | 0.76 | 1.09 |
| Trentino Alto Adige | NE | 1.1 | 14 | 25 | | | | 1.82 | | | | 0.55 | | | | | 0.79 | 5.30 | 0.92 | | | | | | 2.59 |
| Umbria | C | 0.9 | 119 | 95 | 0.30 | 0.05 | 0.47 | 0.74 | 0.98 | 1.02 | 0.84 | 0.94 | 0.32 | 0.38 | 0.16 | 1.42 | 1.02 | 0.37 | 0.90 | 0.38 | 0.72 | 8.05 | | 0.13 | 1.02 |
| Veneto | NE | 4.9 | 328 | 100 | 1.17 | 6.77 | 0.06 | 1.62 | 1.31 | 0.72 | 0.94 | 0.75 | 1.12 | 1.39 | 1.17 | 1.03 | 1.21 | 0.88 | 1.07 | 3.34 | 1.59 | 0.88 | 1.13 | 2.02 | 1.27 |
| | Regional coverage (%) | | | | 80 | 85 | 75 | 90 | 85 | 85 | 90 | 90 | 85 | 70 | 80 | 85 | 90 | 80 | 90 | 75 | 85 | 80 | 60 | 85 | 100 |

** NW, northwest; NE, northeast; C, Center; S, south; I, islands*

*Valle d'Aosta is not listed because the region lacks professors in the specialties under observation.*

*Shaded scores indicate the specialties in which the territory has less than five professors.*

### 4.3 Territorial inequities in knowledge capital and professors' performance

In this section we integrate the analyses of the two previous sections, providing an overall view, for each region, of the research performance (efficiency of production) and knowledge capital (level of research impact) in the specialties on the front lines against the epidemic. For reasons of space we report only the example of Virology. Figure 4 shows the dispersion diagram of FSS and $KC_{PC}$ scores for each region. We recall that the evaluation of research performance includes only university professors, not the researchers of private companies or other institutions.

To facilitate the reading, we divide x-y space in quadrants along the lines of the national averages (i.e. FSS = 1, $KC_{PC}$ = 1). The regions are broadly dispersed over all four quadrants; to the top right (both FSS and $KC_{PC}$ above national average) are Lombardy and Abruzzo, with Piedmont, Umbria and Emilia Romagna also in this quadrant, entering just above the FSS cutoff. In the upper left quadrant (FSS above national average, but $KC_{PC}$ below) we find Liguria, Veneto and Tuscany, with the latter at the very top of the FSS ranking for the specialty. In the lower right quadrant are Trentino Alto Adige, Lazio and Friuli Venezia Giulia ($KC_{PC}$ above national average, but FSS below). These last two regions score highest in Italy by $KC_{PC}$ greater than double the national average, but below the national average by FSS. Finally, in the lower left quadrant (both FSS and $KC_{PC}$ below national average) are seven regions, all from the south and islands except Marche. This positioning shows a gap of southern macro-region with respect to the rest of Italy, for this specialty, both in knowledge capital and academic performance.

***Figure 3: Overview of the Italian regions: normalized knowledge capital per capita ($KC_{PC}$) and average professors' performance (FSS), in Virology (1.10 means 10% above average)***

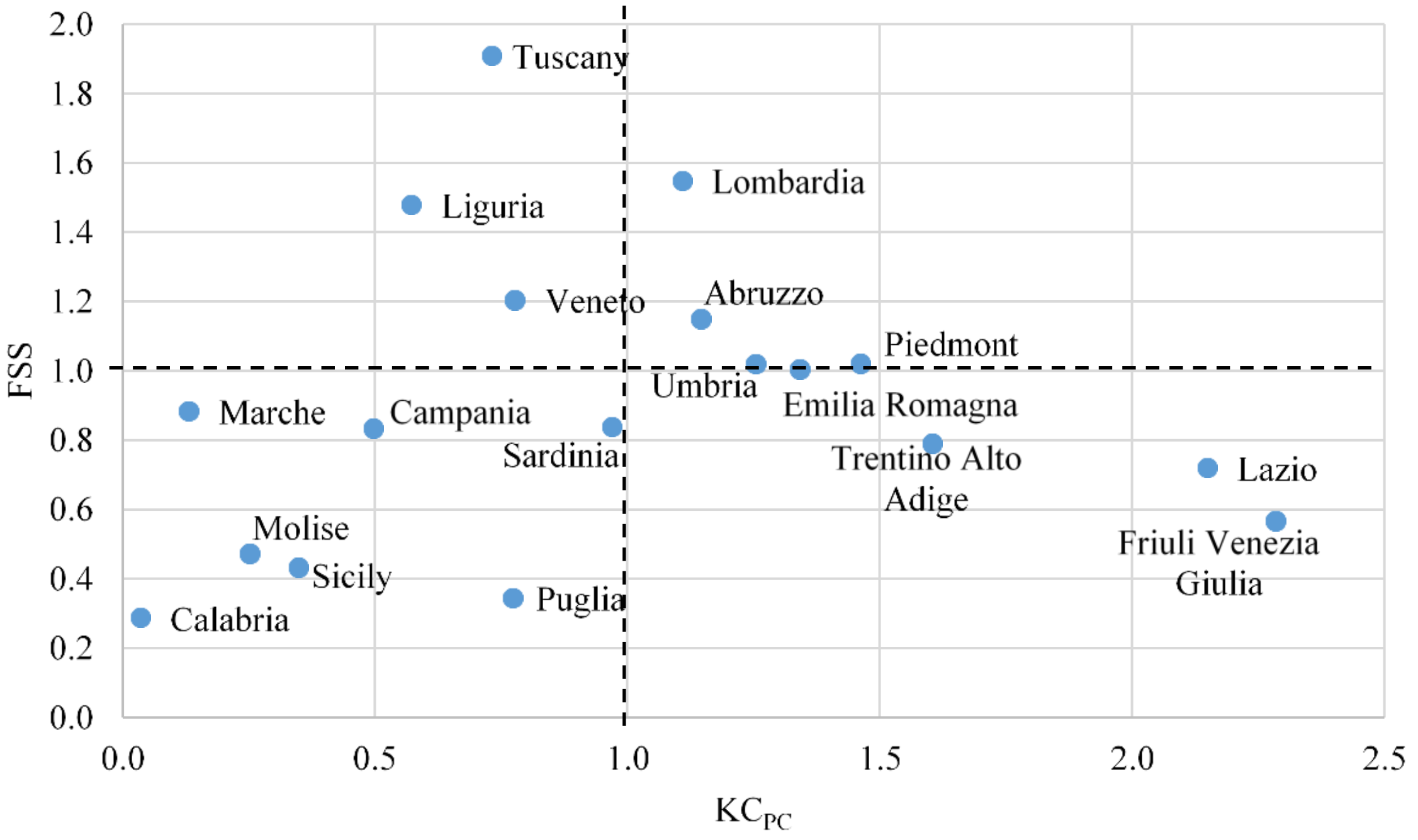

*Basilicata and Valle d'Aosta have no professors in Virology, so are not included.*

The diagramming of Figure 3 could be repeated for any other specialty, for purposes of assessing territorial inequities in both knowledge capital and professors' performance.

In Figure 4 we present a summary of the overall analysis, obtained as an average of the scores detected in the 20 specialties. The diagram does not show Trentino Alto Adige ($KC_{PC}$ = 0.59), an extreme outlier for FSS (2.6).

In the upper right quadrant we find one central region (Tuscany) and four of the northern regions that bore the heaviest brunt of the virus (Lombardy, Emilia Romagna, Veneto and Liguria). However, Piedmont, another northern region heavily impacted, is positioned in the lower left quadrant, together with Marche (center) and all the regions of the islands and south, but Abruzzo, which in terms of $KC_{PC}$ shows a score of 1.05 (5% above national average), arriving just inside the lower right quadrant, where there are also Friuli Venezia Giulia (north) and Lazio (center). The latter records the maximum score of $KC_{PC}$ (1.71) against an FSS more than 25% lower than the national average. Two regions place in the upper left quadrant (FSS above national average and $KC_{PC}$ below), Trentino Alto Adige (north), out of scale for FSS, and Umbria (center).

***Figure 4: Overview of the Italian regions: normalized knowledge capital per capita ($KC_{PC}$) and average professors' performance (FSS) in the 20 COVID-19 front-line specialties (1.10 means 10% above average)***

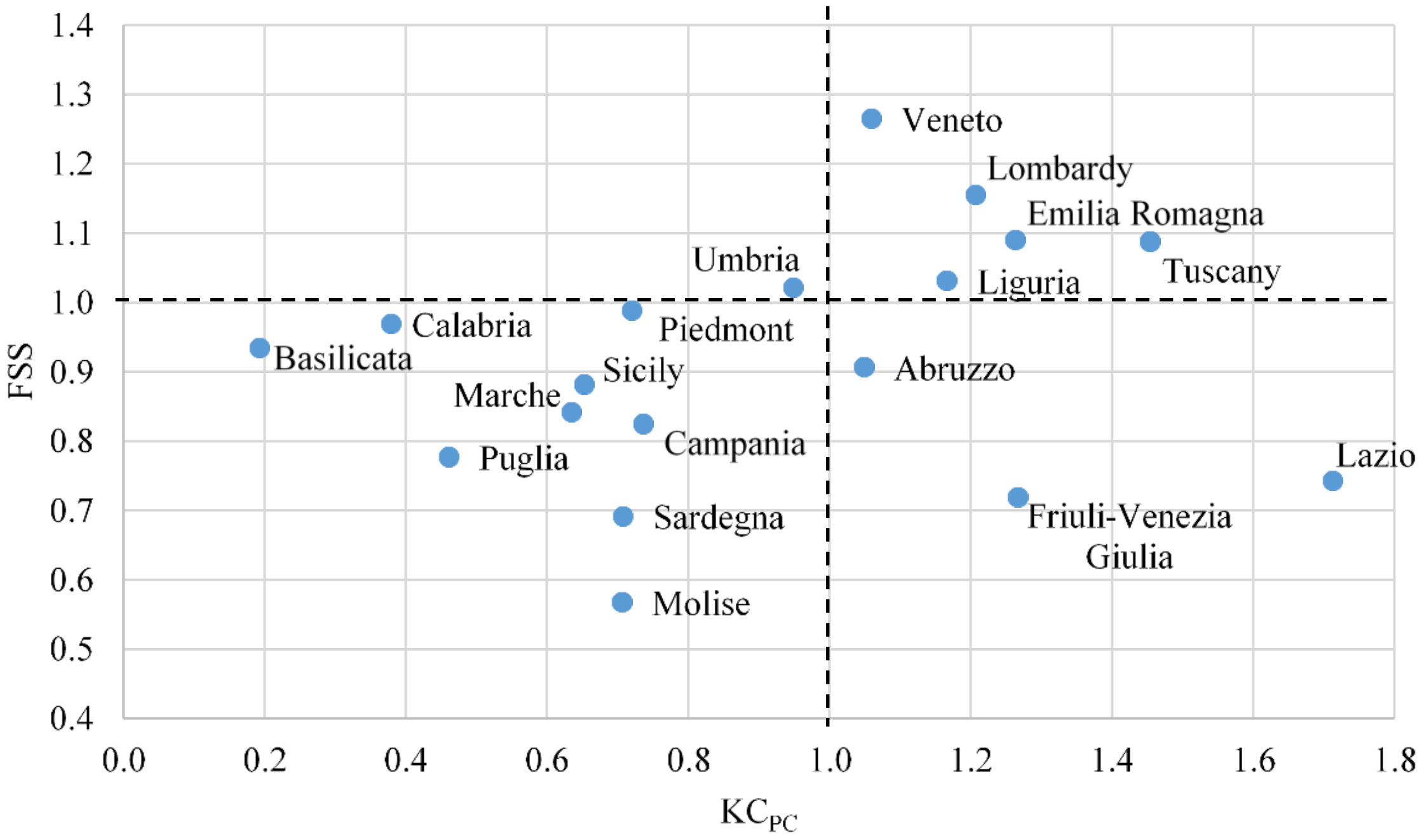

*Valle d'Aosta has no professors in the 20 front-line specialties, so is not included.*
*Trentino Alto Adige is out of scale for FSS (2.6)*

Figure 5 presents the same analysis for the Italian macro-regions. The northwest and northeast stand out for both productivity of their university professors and the KC produced per capita. The south and islands macro-regions, instead show much lower scores in the positioning of their research systems. The territories of the center achieve the highest $KC_{PC}$ score, but suffer a gap in the performance of their professors compared to the north.

*Figure 5: The Italian macro-regions: knowledge capital per capita ($KC_{PC}$) and average professors' performance (FSS) in the 20 specialties*

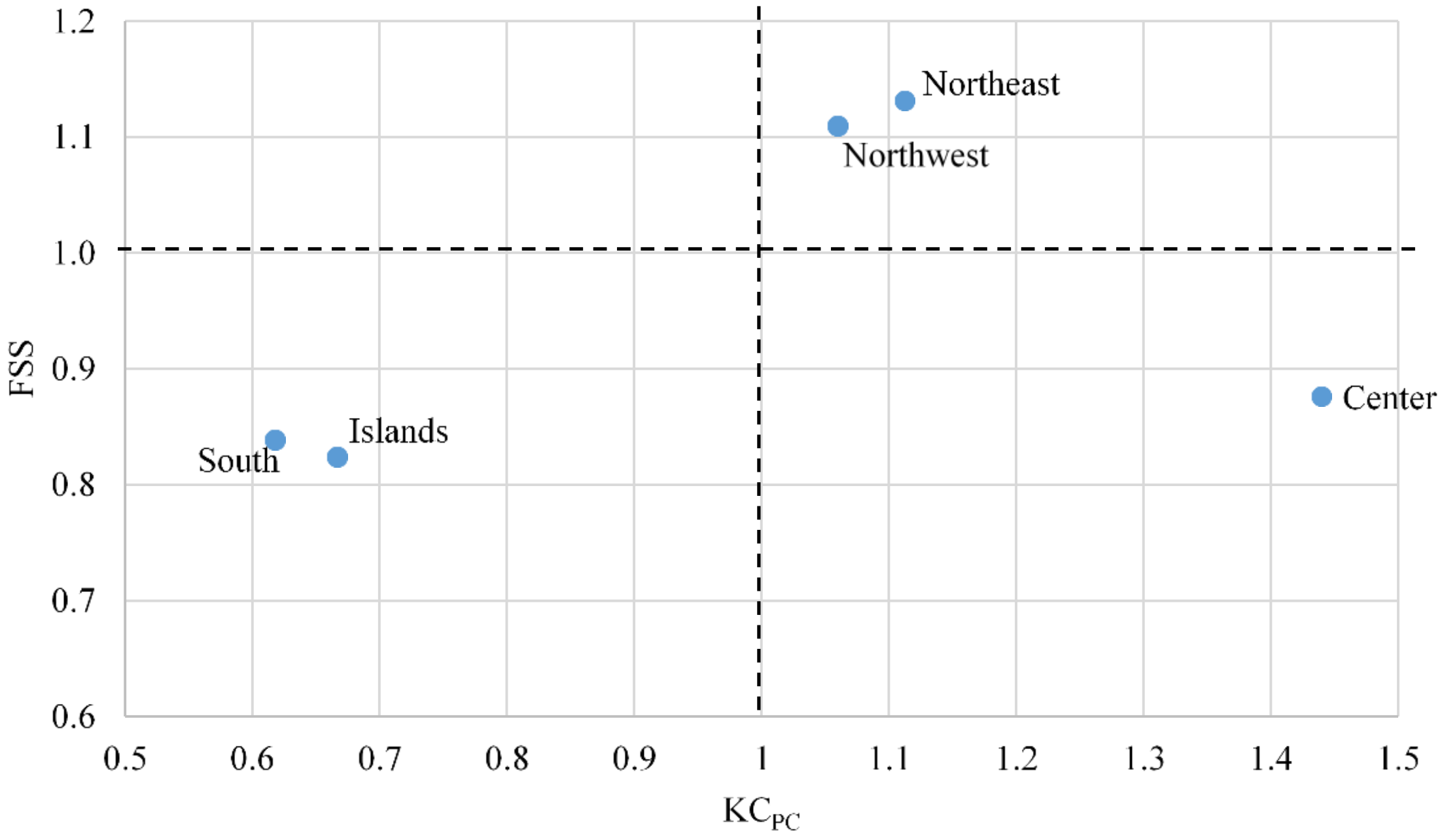

## 5. Conclusions

Measuring the intangible assets of an organization or a territory is a formidable task. However, thanks to the advances of scientometrics (notwithstanding its limits and embedded assumptions), we can now quantify and value scientific knowledge.

In this article, using quantitative metrics of "knowledge capital" and "performance of professors", we have developed a methodology to assess the scientific wealth of territories, and applied it to map its distribution in Italy in 20 medical specialties engaged on the front lines against COVID-19. The analysis can be easily expanded to all medical specialties, to the STEM and economic sciences, and replicated in other countries.

In the current global context, knowledge is the key driver of socio-economic development, making this kind of scientometrics application particularly important. Indeed, socio-economic policies should proceed only following strategic analysis of the country's scientific wealth, including its sectoral and territorial distribution. However, the territorial inequities in national and regional health systems have thus far been analyzed essentially in terms of tangible assets, such as capacities in hospital beds, equipment, numbers of personnel and the like. With the application of scientometrics techniques, national and regional decision-makers can now also draw on the assessment of scientific wealth, in the formulation and monitoring of policies.

The results of our analysis of the strengths and weaknesses in the medical sciences, in each Italian province, region and macro-region, confirm that in terms of the capacities to respond to an event such as the COVID-19 pandemic, the country once again suffers from the classic north-south divide in socio-economic development. The disadvantage of southern healthcare was already widely known and reported in terms of tangible assets (SVIMEZ, 2019), but now we see that this also clearly extends to the scientific wealth underlying the capacities and quality of service.

Four northern regions (two northwest, two northeast) and one in the center score above national average in both knowledge capital per capita and professors' research performance. In contrast, apart from Abruzzo, all southern and island regions lag below average in both indicators. We recall that Italian professors in the medical sciences serve in numerous roles, as researchers, but also educators, practitioners, consultants, and that they have been called to serve in the COVID-19 advisory task forces for national and regional governments. Given these multiple roles, the assessment of their performance should be extremely relevant.

Our findings support the observations communicated by authoritative physicians and local governors, that if the pandemic had struck first in the south of Italy there would have been even worse consequences. These observations are accompanied by warnings concerning the need for quick intervention to rebalance the territorial inequities in health care. As our study shows, the policies for rebalancing the north-south divide should also consider, in addition to tangible assets, the gap in production and availability of quality medical knowledge. The national government could well consider these inequities in establishing the criteria for allocating the important financial resources recently identified for health research, thanks in part to the support of the EU.

In concluding this study, we recall the main limits embedded in bibliometric analyses. First, the new knowledge produced is not only that embedded in publications, and the bibliographic repertories (such as WoS, used here) do not register all publications. Second, the measurement of the value of publications using citation-based indicators is a prediction, not definitive. Citations can also be negative or inappropriate, and in any case they certify only scholarly impact, forgoing other types of impact. Third, given the limitations on input data (e.g. costs of labor and capital), we make several assumptions in measuring professors' performance. Finally, the results could be sensitive to the classification schemes used for publications and professors, and to the number and type of specialties chosen for analysis. Such limitations should induce caution in interpreting any findings arising from scientometrics techniques, however, as always, they should not be systemic to any specific territory. In this case, given the amplitude of observations in the fields under analysis, we are confident that the reality is quite similar to the picture produced here.